\documentclass[12pt]{iopart}

\usepackage{graphicx}
\usepackage{amstext}

\newcommand {\pT}       {\ensuremath{p_{\mathrm T}}}
\newcommand {\PbPb}  {\mbox{PbPb}}
\newcommand {\GeVc} {\mbox{GeV/c}}
\newcommand {\MeVc} {\mbox{MeV/c}}
\newcommand {\pp}    {\mbox{pp}}
\newcommand {\AJ}       {\ensuremath{A_J}}

\begin{document}

% Usage: \title[Short title]{Full title}
% [Short title] is optional; use where title is too long
% or contains footnotes, 50 characters maximum 
\title{Studies of Jet Quenching in PbPb collisions at CMS}

\author{Matthew Nguyen for the CMS Collaboration}

\address{CERN, Meyrin, Switzerland}
\ead{Matthew.Nguyen@cern.ch}

\begin{abstract}
Jets are an important tool to probe the hot, dense medium produced in ultra-relativistic heavy-ion collisions.  At the collision energies available at the Large Hadron Collider (LHC), there is copious production of hard processes, such that high \pT\ jets may be differentiated from  the heavy-ion underlying event. The multipurpose Compact Muon Solenoid (CMS) detector is well designed to measure hard scattering processes with its high quality calorimeters and high precision silicon tracker~\cite{cms}. Jet quenching has been studied in CMS in PbPb collisions at $\sqrt{s_{_{NN}}}=$\,2.76~TeV. As a function of centrality, dijet events with a high \pT\ leading jet were found to have an increasing momentum imbalance that was significantly larger than predicted by simulations. The angular distribution of jet fragmentation products has been explored by associating charged tracks with the jets measured in the calorimeters.  By projecting the momenta of charged tracks onto the leading jet axis it is shown that the apparent momentum imbalance of the leading dijet pair can be recovered if low \pT\ tracks are considered.  A large fraction of the balancing momentum carried by these soft particles is radiated at large angle relative to the jets. 

\end{abstract}

Jets associated with the hard scattering of partons are a powerful probe of the hot, dense matter created in heavy-ion collisions, which is believed to be a Quark-Gluon Plasma (QGP).  Previous data, mostly in the form of single and di-hadron observables show that jets are strongly modified by the medium, a phenomenon known as jet quenching~\cite{dEnterria}. The large \PbPb\ collision energies at the LHC provide an abundance of jets of \pT\ $>$ 100 \GeVc, facilitating the direct reconstruction of jets.  We review recent studies of jet quenching in \PbPb\ collisions at a nucleon-nucleon center-of-mass energy of $\sqrt{s_{_{NN}}}=2.76$~TeV collected in 2010 using the Compact Muon Solenoid (CMS) detector.  The results presented in these proceedings are a subset of those found in~\cite{ourPaper}.

% Advertise at the end?
%Further studies may be found in~\cite{refOurPaper}.
%A total integrated luminosity of 8.7~$\mu$b$^{-1}$ was collected, of which 6.7~$\mu$b$^{-1}$ is included in this analysis.                                                                                                                  

Jets are reconstructed from the energy deposited in the lead-tungstate crystal electromagnetic
calorimeter (ECAL) and the brass/scintillator hadron calorimeter (HCAL) covering $|\eta| < 3$.
The steel/quartz-fiber Cherenkov Hadron Forward (HF) calorimeter, covering
$3<|\eta|<5.2$ is  used for centrality determination.
Calorimeter cells are grouped into towers of granularity $\Delta \eta \times \Delta \varphi =0.087\times0.087$ in the barrel region ($|\eta| < 1.5$), and a somewhat coarser segmentation in the endcaps.

The CMS tracking system, located inside the calorimeters, consists of pixel and silicon-strip layers covering  $|\eta| <                       
2.5$. It provides track reconstruction down to $\pT \approx 100$~\MeVc, with
a track momentum resolution of about 1\% at $\pT = 100$~\GeVc. The tracking system and central calorimeters are embedded in a solenoid with 3.8~T central magnetic field.
A set of scintillator tiles on the inner side of the HF calorimeters, the Beam Scintillator Counters (BSC), provide triggering and beam-halo rejection.
%\cite{bib_geant}.

Jet events were selected using the High Level Trigger, requiring a jet  with  $\pT > 50$~\GeVc, where the jet $\pT$ value is uncorrected 
for the calorimeter response. The trigger becomes fully efficient for collisions with a leading jet
with corrected $\pT$ greater than 100~\GeVc. The large underlying event in heavy-ion collisions is subtracted on an event-by-event basis according to the procedure described in~\cite{Kodolova:2007hd}.
Jets are reconstructed using an iterative cone algorithm~\cite{cone}.
Jet corrections for the calorimeter response have been applied, as determined in studies for \pp\
collisions~\cite{CMS-PAS-JME-10-010}. 

To obtain a pure sample of dijets the following selection was applied:
\begin{itemize}
\item Leading jet:  corrected jet $p_{\mathrm{T},1} > 120$~\GeVc\ and $|\eta_1| <  2$
\item Subleading jet:  corrected jet $p_{\mathrm{T},2} > 50$~\GeVc\ and $|\eta_2| <  2$
\item Azimuthal angle between the jets of $\Delta\phi_{12}>2\pi/3$ radians
\end{itemize}

Prior to jet finding on the selected events, a small contamination of
noise events from uncharacteristic ECAL and HCAL detector responses was removed using signal
timing, energy distribution, and pulse-shape information \cite{noise}. As a result,
about 2.4\% of the events were removed from the sample.

As a baseline for quenching effects the data are compared to dijets in {\sc{pythia}}, representing a sample with no quenching.
To simulate the effects of the heavy-ion underlying event these dijet events are embedded into both real \PbPb\  data and simulated \PbPb\ data using the {\sc {hydjet}} generator~\cite{Lokhtin:2005px}.  Both embedded samples were
propagated through the standard reconstruction and analysis chain.

\begin{figure}[th!]
\begin{center}
\includegraphics[width=0.9\textwidth]{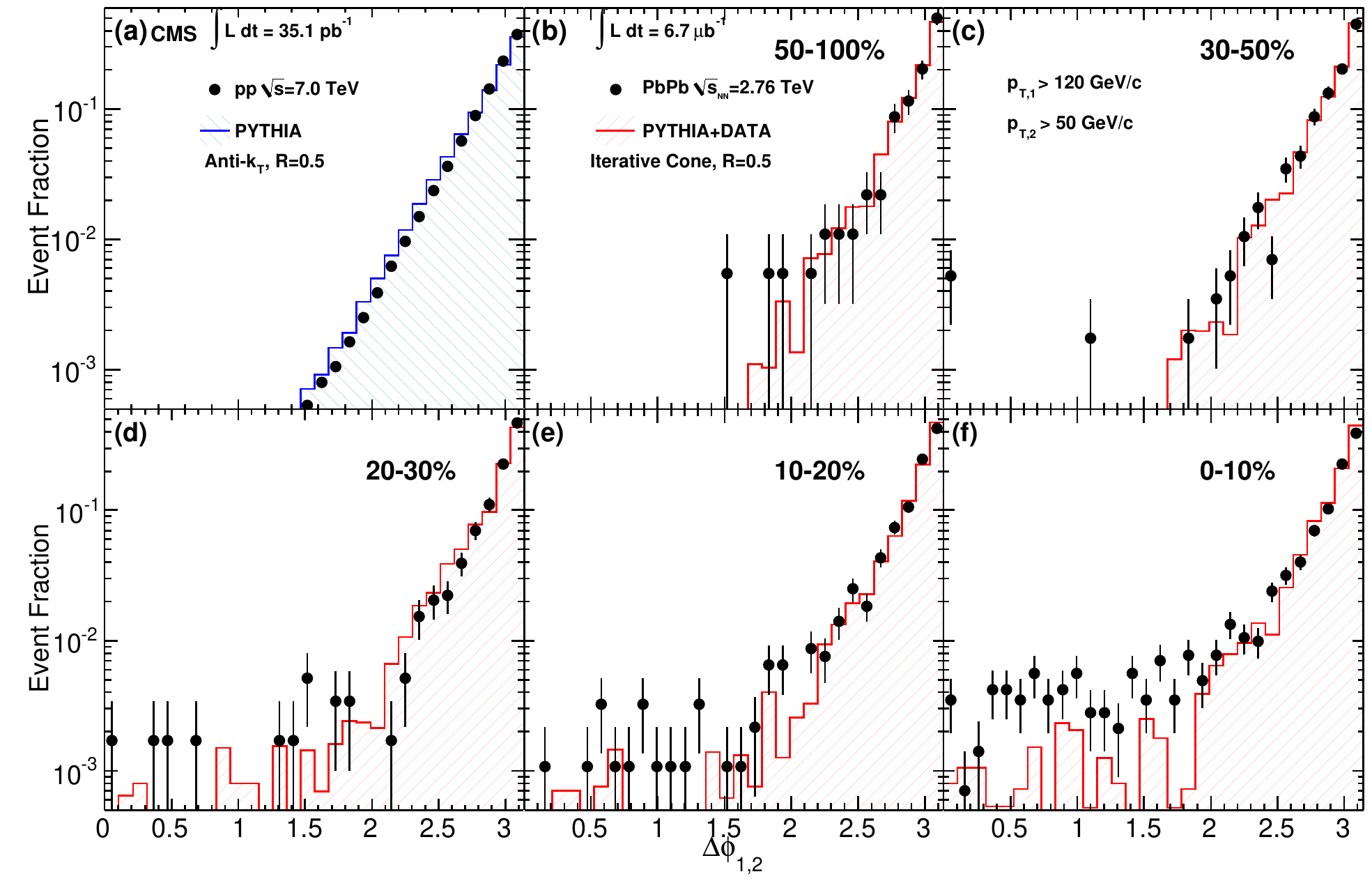}
\caption{$\Delta \phi_{12}$ distributions for leading jets of $p_{\mathrm{T},1} > 120$~\GeVc\
with subleading jets of $p_{\mathrm{T},2} >  50$~\GeVc\ for
7 TeV pp collisions (a) and 2.76 TeV PbPb collisions in several centrality bins:
(b) 50--100\%, (c) 30--50\%, (d) 20--30\%, (e) 10--20\% and (f) 0--10\%.
Data are shown as points, while the histograms show (a) {\sc{pythia}} events  and (b)-(f) {\sc{pythia}} events embedded into \PbPb\ data.
The error bars show the statistical uncertainties.}
\label{fig:dphi50}
\end{center}
\end{figure}

Figure~\ref{fig:dphi50} shows distributions of $\Delta \phi_{12}$ between leading and subleading jets which pass the respective \pT\ selections.   
Figure~\ref{fig:dphi50} (a) shows \pp\ data at 7 TeV compared to {\sc{pythia}} simulations, while Fig.~\ref{fig:dphi50} (b)-(f) show \PbPb\ data in five centrality bins,
compared to {\sc{pythia+data}} simulations.  In general, the distributions agree quite well with the {\sc{pythia}} reference simulations.
The more central events show an excess of events with azimuthally misaligned dijets
($\Delta \phi _{12} < 2$), compared with more peripheral events.  A similar trend is seen for the {\sc{pythia+data}} simulations, although the
fraction of events with  azimuthally misaligned dijets is smaller in the simulation. 
The tails of these distributions in central events can
be understood as the result of the increasing rate of fake jets or mismatched jets which come from another hard scattering.  The effect is larger in data than in simulation, as the subleading jet can undergo a sufficiently large energy loss to fall below the 50~\GeVc\
selection criteria.

\begin{figure}[th!]
\begin{center}
\includegraphics[width=0.9\textwidth]{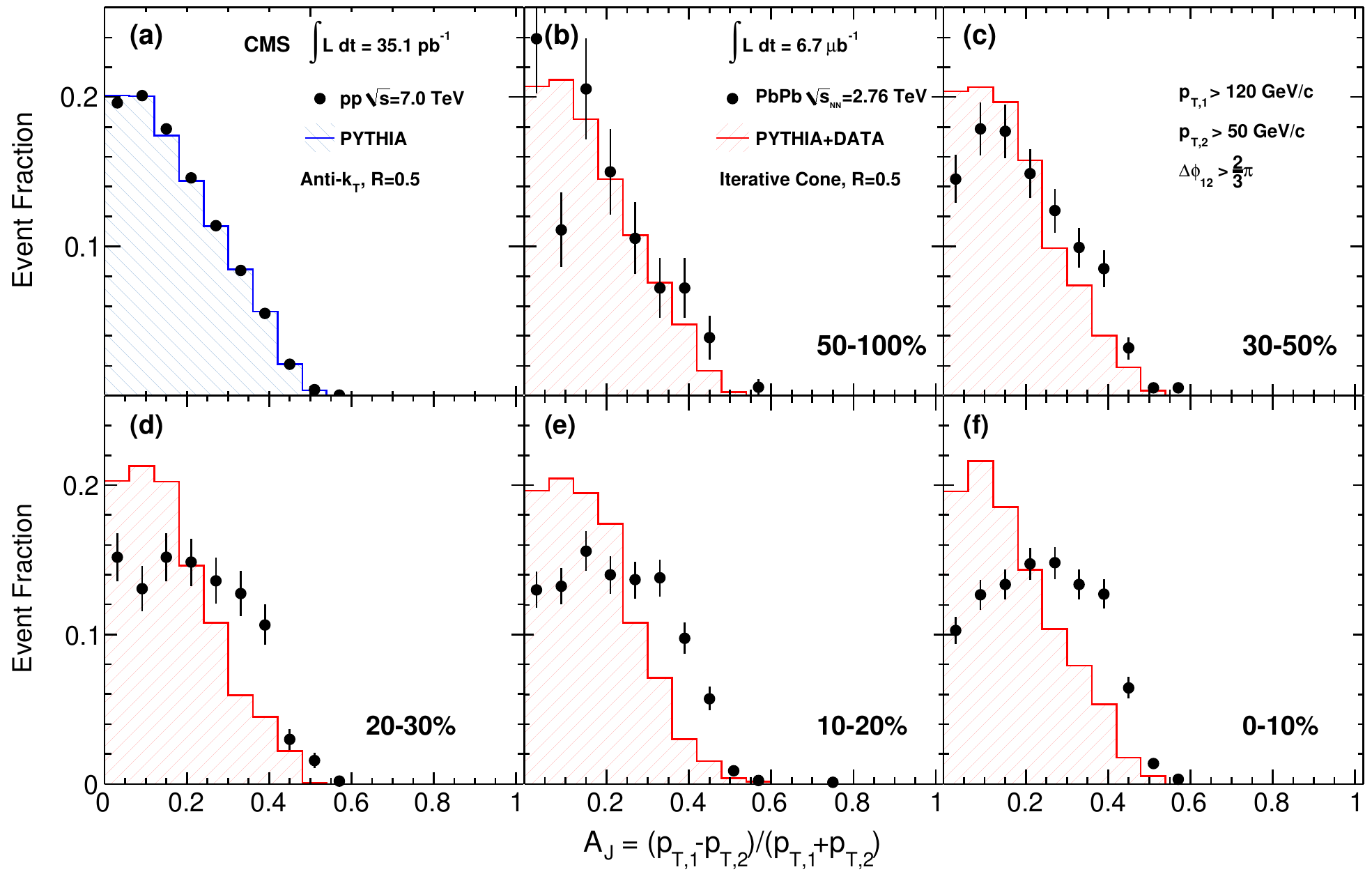}
\caption{Dijet asymmetry ratio, $A_{J}$, for leading jets of $p_{\mathrm{T},1}> $ 120~\GeVc, subleading jets of $p_{\mathrm{T},2}> $50~\GeVc\, and $\Delta\phi_{12}>2\pi/3$  for 7 TeV pp collisions (a) and 2.76 TeV \PbPb\ collisions in several centrality bins:  (b) 50--100\%, (c) 30--50\%, (d) 20--30\%, (e) 10--20\% and (f) 0--10\%.
Data are shown as points, while the histograms show (a) {\sc{pythia}} events and  (b)-(f) {\sc{pythia}} events embedded into \PbPb\ data.
The error bars show the statistical uncertainties.}
\label{fig:JetAsymm}
\end{center}
\end{figure}

To characterize the \pT\ balance of the dijet, the asymmetry variable \AJ\ is used, where
$A_J \equiv ({p_{\mathrm{T},1}-p_{\mathrm{T},2}})/({p_{\mathrm{T},1}+p_{\mathrm{T},2}})$.
The \AJ\ distribution for \pp\ collisions at 7 TeV, plotted In Fig.~\ref{fig:JetAsymm} for \pp\ (a), agrees well with {\sc{pythia}}.
The centrality dependence of $A_J$ for \PbPb\ collisions can be seen in
Figs.~\ref{fig:JetAsymm}~(b)-(f), in comparison to {\sc{pythia+data}} simulations.
Whereas the dijet angular correlations show only a small dependence on collision
centrality, the dijet momentum balance exhibits a dramatic change in shape
for the most central collisions. In contrast, the {\sc{pythia}} simulations only
exhibit a modest broadening, even when embedded in the highest multiplicity
\PbPb\ events.  The large rate of highly imbalanced jets indicates a strong jet quenching effect in which energy no longer reaches the calorimeters inside the jet cone. 
The absence of any modification to the $\Delta \phi_{12}$ distributions suggests that this energy is not transferred via hard radiation.

To fate of the ``lost'' energy was studied in greater detail by looking at correlations of the jets with charged tracks.
The distribution of jet-associated tracks was studied as a function of both track \pT\ and $\Delta R$ from the leading and subleading jet axis (not shown).
The background of combinatorial jet-track pairs was explicitly subtracted.  It was found, however, that the size of the background limited the study
to tracks with $p_{\mathrm{T}} > 1.0$~\GeVc\  and $\Delta R < 0.8$.  To pursue the fate of the lost energy outside of this domain, a more inclusive quantity was studied.
The  overall momentum balance in the dijet events can be obtained using the projection of missing
\pT\ of reconstructed charged tracks onto the leading jet axis. For each event, this
projection was calculated as
%for each event with a leading jet with $p_{\mathrm{T},1} > 120$~\GeVc\ and $|\eta_1| < 2$:                                                   
\begin{equation*}
\displaystyle{\not} p_{\mathrm{T}}^{\parallel} =
\sum_{\rm i}{ -p_{\mathrm{T}}^{\rm i}\cos{(\phi_{\rm i}-\phi_{\rm Leading\ Jet})}},
%\end{eqnarray*}                                                                                                                              
\end{equation*}
where the sum is evaluated over all tracks with $\pT > 0.5$~\GeVc\ and $|\eta| < 2.4$. The results were
then averaged over events to obtain $\langle \displaystyle{\not} p_{\mathrm{T}}^{\parallel} \rangle$.
No explicit background subtraction is applied in this method, as the heavy-ion underlying event is not expected to give a net \pT\ contribution along the leading jet axis.

\begin{figure}[th!]
  \begin{center}
    \includegraphics[width=0.8\textwidth]{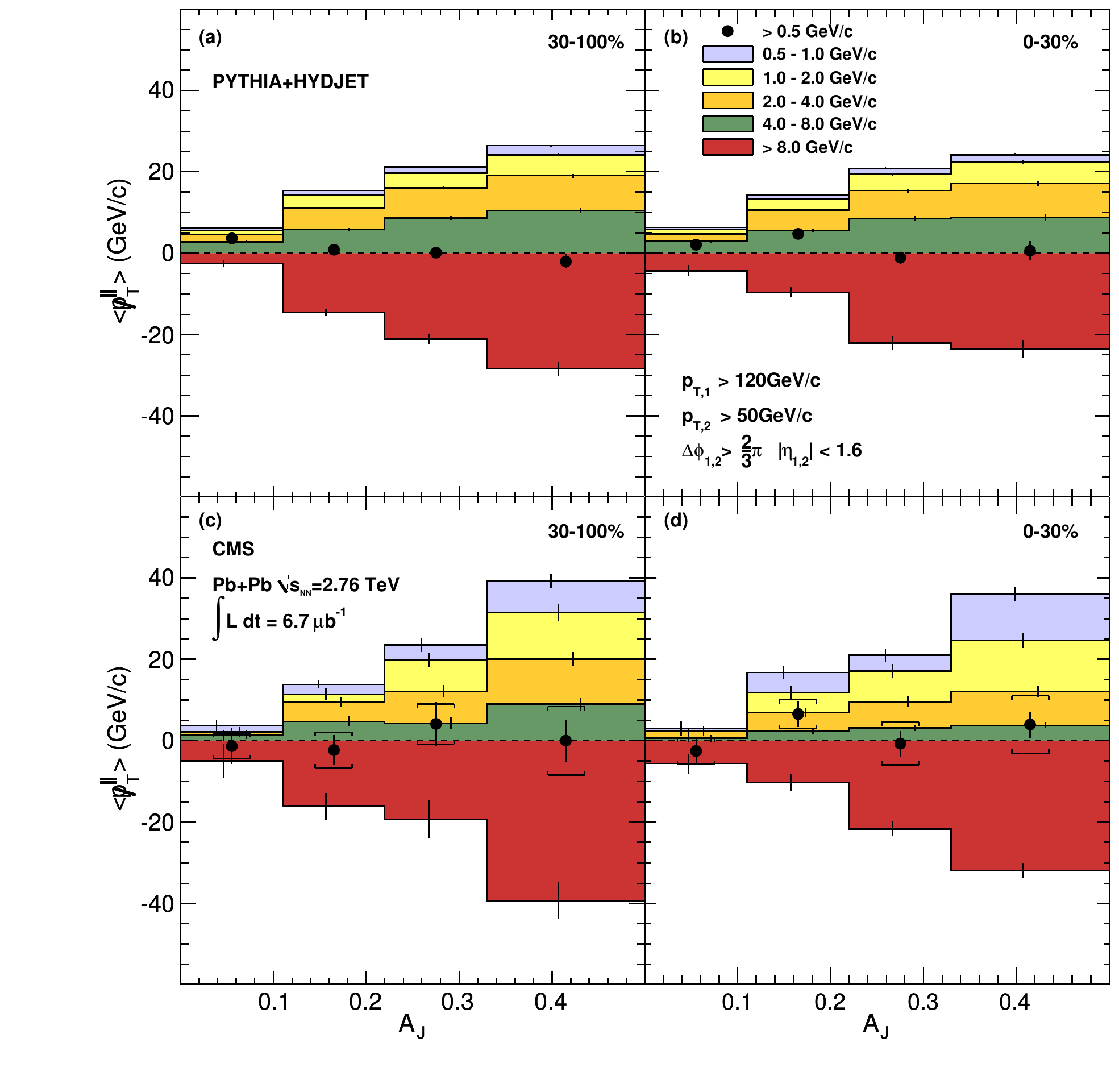}
    \caption{Average missing transverse momentum,
$\langle \displaystyle{\not} p_{\mathrm{T}}^{\parallel} \rangle$,
for tracks with $\pT > 0.5$~${\rm GeV/c} $, projected onto the leading jet axis (solid circles).
The $\langle \displaystyle{\not} p_{\mathrm{T}}^{\parallel} \rangle$ values are shown as a function of dijet asymmetry
$A_J$ for 30--100\% centrality (left) and 0--30\% centrality (right).
For the solid circles, vertical bars and brackets represent
the statistical and systematic uncertainties, respectively.
Colored bands show the contribution to $\langle \displaystyle{\not} p_{\mathrm{T}}^{\parallel} \rangle$ for five
ranges of track \pT. The top and bottom rows show results for {\sc{pythia+hydjet}} and \PbPb\ data, respectively.
For the individual \pT\ ranges, the statistical uncertainties are shown as vertical bars.
}
    \label{fig:MissingpT}
  \end{center}
\end{figure}

%\begin{eqnarray*}                                                                                                                            

In Fig.~\ref{fig:MissingpT}, the solid markers show $\langle \displaystyle{\not} p_{\mathrm{T}}^{\parallel} \rangle$
as a function of \AJ\ for two centrality bins, 30--100\% (left) and 0--30\% (right).
The top row shows simulation ({\sc{pythia}} embedded in {\sc{hydjet}}), while the bottom row shows \PbPb\ data.
Even for large \AJ\ dijet events, the charged tracks above $\pT > 0.5$~\GeVc\
show no net momentum balance with respect to the leading jet axis in data or simulation.
The figure also shows the contributions to $\langle \displaystyle{\not} p_{\mathrm{T}}^{\parallel} \rangle$ for five transverse momentum ranges from 0.5--1~\GeVc\ to
$\pT > 8$~\GeVc. The vertical bars for each range denote statistical uncertainties.
 For data and simulation, a large negative contribution to $\langle \displaystyle{\not} p_{\mathrm{T}}^{\parallel} \rangle$
(i.e., in the direction of the leading jet) by the $\pT > 8$~\GeVc\ range is balanced by the combined contributions from
the  0.5--8~\GeVc\ regions.  Looking at the $\pT < 8$~\GeVc\ region in detail, important differences
between data and simulation emerge. For {\sc{pythia+hydjet}} both centrality ranges show a large
balancing contribution from the intermediate \pT\ region of 4--8~\GeVc, while the contribution from
the two regions spanning 0.5--2~\GeVc\ is very small. In peripheral \PbPb\ data, the contribution of
0.5--2~\GeVc\ tracks relative to that from 4--8~\GeVc\ tracks is somewhat enhanced compared to the simulation.
In central \PbPb\ events, the relative contribution of low and intermediate-\pT\  tracks
is actually the opposite of that seen in {\sc{pythia+hydjet}}.
In data, the 4--8~\GeVc\ region makes almost no contribution to the overall
momentum balance, while a large fraction of the negative imbalance from high \pT\ is recovered in low-momentum tracks.

\begin{figure}[th!]
  \begin{center}
    \includegraphics[width=0.8\textwidth]{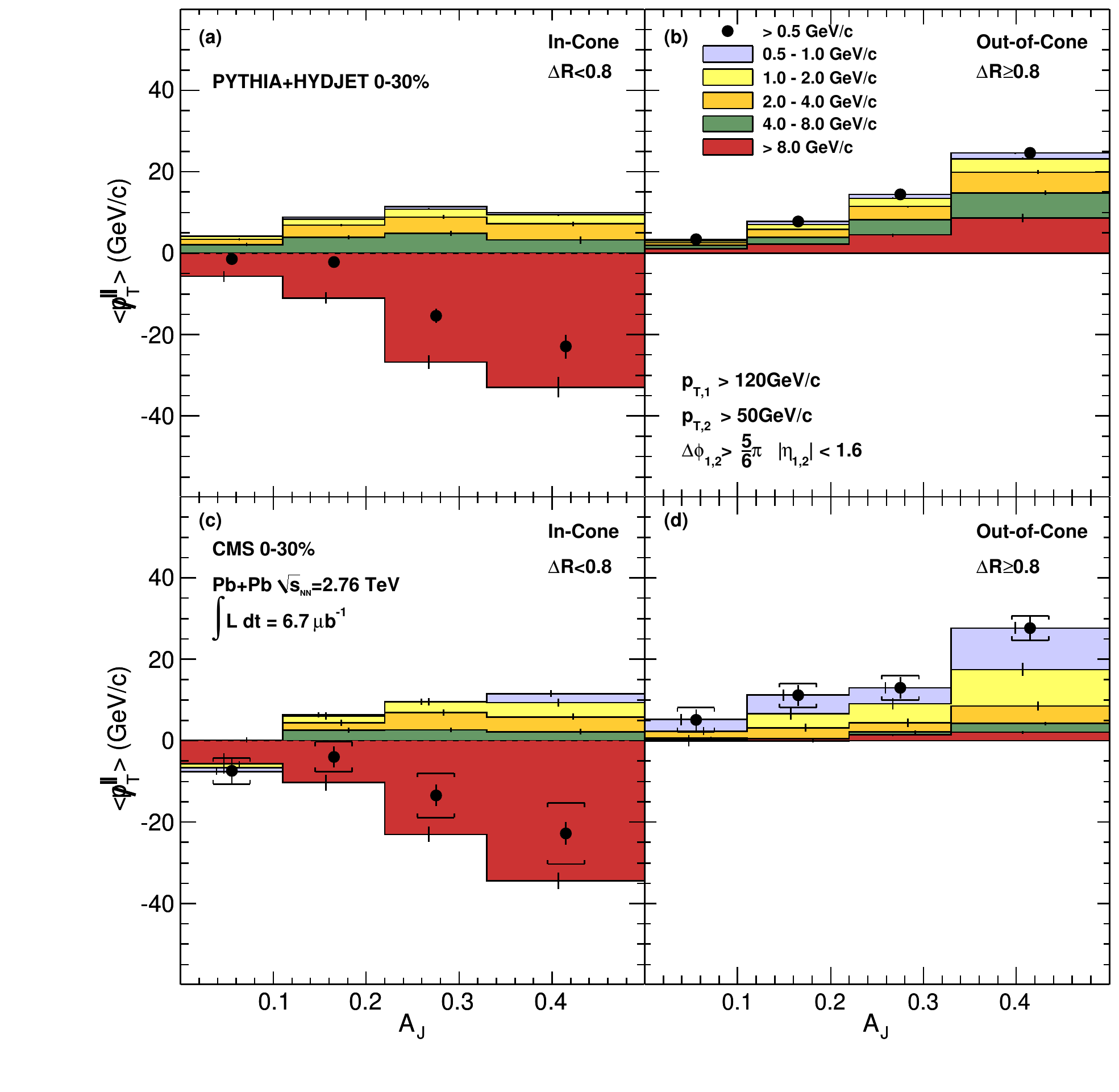}
    \caption{Average missing transverse momentum,
$\langle \displaystyle{\not} p_{\mathrm{T}}^{\parallel} \rangle$,
for tracks with $\pT > 0.5 {\rm GeV/c}$, projected onto the leading jet axis (solid circles).
The $\langle \displaystyle{\not} p_{\mathrm{T}}^{\parallel} \rangle$ values are shown as a function of dijet asymmetry
$A_J$ for 0--30\% centrality, inside ($\Delta R < 0.8$) one of the leading or subleading jet cones (left) and
outside ($\Delta R > 0.8$) the leading and subleading jet cones (right).
For the solid circles, vertical bars and brackets represent
the statistical and systematic uncertainties, respectively.
For the individual \pT\ ranges, the statistical uncertainties are shown as vertical bars.  }
    \label{fig:MissingpTInConeOutCone}
  \end{center}
\end{figure}

Further insight into the radial dependence of the momentum balance can be gained by studying
$\langle \displaystyle{\not} p_{\mathrm{T}}^{\parallel} \rangle$ separately for tracks inside cones of size $\Delta R = 0.8$
around the leading and subleading jet axes, and for tracks outside of these cones. The results of this study for central events
are shown in Fig.~\ref{fig:MissingpTInConeOutCone} for the in-cone balance and out-of-cone balance for MC and data.
%As the underlying \PbPb\ event in both data and MC is not $\phi$-symmetric on an event-by-event basis, the back-to-back requirement was tightened to $\Delta \phi_{12} > 5 \pi/6$ for this study.
One observes that for both data and simulation an in-cone imbalance of $\langle \displaystyle{\not} p_{\mathrm{T}}^{\parallel} \rangle \approx        
-20$~\GeVc\ is found for the $\AJ > 0.33$ selection. In both cases this is balanced by a corresponding out-of-cone
imbalance of  $\langle \displaystyle{\not} p_{\mathrm{T}}^{\parallel} \rangle \approx 20$~\GeVc.  However in simulation, more than 50\% of the balance is carried by tracks with 
$\pT > 4$~\GeVc, as might be expected from multijet production, whereas in data the balance is carried almost entirely by tracks with $0.5 < \pT < 4$~\GeVc.

To conclude, detailed studies of dijet in \PbPb\ collisions have been performed.
The increase in the frequency of imbalanced dijets in central events demonstrates that a sizeable amount of energy is transferred out of the subleading jet.
The absence of any strong angular decorrelation of the jet pairs disfavors a scenario in which this energy takes the form of hard radiation.  By examining the \pT\ balance of charged tracks with respect to the leading dijet axis, it was found that asymmetric dijet events are balanced when tracks are considered down to \pT\ of 500 \MeVc.  In \PbPb\ data this balance is dominated by tracks below $\sim$ 2 \GeVc.  By studying the balance of charged tracks both inside and outside a cone of R = 0.8, it was found that the majority of this energy is distributed at large angle with respect to the jet axis.

\end{document}